\documentstyle[twoside,fleqn,espcrc2]{article}

\batchmode
  \newfont{\bbbfont}{msbm10}
\errorstopmode

\newif\ifamsf\amsftrue
\ifx\bbbfont\nullfont
  \amsffalse
\fi

\newfont{\smallbbbfont}{msbm7}
\newfont{\tinybbbfont}{msbm5}
\newfont{\footbbbfont}{msbm8}
\newfont{\smallfootbbbfont}{msbm6}

\newif\iffn\fnfalse

\ifamsf
  \newcommand{\Bbb}[1]{\iffn
      \mathchoice{\mbox{\footbbbfont #1}}{\mbox{\footbbbfont #1}}
      {\mbox{\smallfootbbbfont #1}}{\mbox{\tinyfootbbbfont #1}}\else
      \mathchoice{\mbox{\bbbfont #1}}{\mbox{\bbbfont #1}}
      {\mbox{\smallbbbfont #1}}{\mbox{\tinybbbfont #1}}\fi}
\else
  
  \def\Bbb{\bf}
\fi

\newcommand{\cE}{{\cal E}}
\newcommand{\cL}{{\cal L}}
\newcommand{\cM}{{\cal M}}
\newcommand{\C}{{\Bbb C}}
\renewcommand{\P}{{\Bbb P}}
\newcommand{\R}{{\Bbb R}}
\newcommand{\Z}{{\Bbb Z}}

\begin{document}
\setcounter{page}0
\thispagestyle{empty}
\hskip5in
\vbox{\hbox{DUKE-TH-95-102}
\hbox{hep-th/9512016}
}

\vskip1in

\begin{center}
{\LARGE Mirror Symmetry and the Type II String}\\

\vskip.3in

{\large David R. Morrison}\\

\vskip.3in

{Department of Mathematics\\
Box 90320\\
Duke University\\
Durham, NC 27708-0320 USA}
\end{center}

\vskip.8in

\begin{center}
{\large\bf Abstract}
\end{center}

{\large
If $X$ and $Y$ are a mirror pair of Calabi--Yau threefolds, mirror
symmetry should extend to an isomorphism between the type IIA string
theory compactified on $X$ and the type IIB string theory compactified
on $Y$, with all nonperturbative effects included.  We study the
implications which this proposal has for the structure of the semiclassical
moduli spaces of the compactified type II theories.  For the type IIB
theory, the form taken
by discrete shifts in the Ramond-Ramond scalars exhibits an unexpected
dependence on the $B$-field.
(Based on a talk at the Trieste Workshop on S-Duality and Mirror Symmetry.)
}

\vfill
\eject

\title{Mirror Symmetry and the Type II String}

\author{David R. Morrison
\address{Department of Mathematics, Box 90320,
Duke University, Durham, NC 27708-0320, USA}
}

\begin{abstract}
If $X$ and $Y$ are a mirror pair of Calabi--Yau threefolds, mirror
symmetry should extend to an isomorphism between the type IIA string
theory compactified on $X$ and the type IIB string theory compactified
on $Y$, with all nonperturbative effects included.  We study the
implications which this proposal has for the structure of the semiclassical
moduli spaces of the compactified type II theories.  For the type IIB
theory, the form taken
by discrete shifts in the Ramond-Ramond scalars exhibits an unexpected
dependence on the $B$-field.
\end{abstract}

\maketitle

\section{INTRODUCTION}

The dramatic recent progress in understanding nonperturbative aspects
of string theory has come about through a study of various
proposed equivalences
between (perturbatively formulated) string theories.  One such
equivalence---one
which has received relatively little attention in this regard---is
mirror symmetry.
Perhaps less attention has been paid  because mirror symmetry is
not a duality relating strong and weak string-couplings.  However, the
hypothesis
that mirror symmetry extends to an equivalence between nonperturbative
string theories has some interesting consequences for those theories
\cite{udual}, which
we will review and extend here.

The equivalence which we consider relates the IIA string theory compactified
to 4 dimensions on a Calabi--Yau manifold $X$, to the IIB theory
compactified on the mirror partner $Y$ of $X$.\footnote{The similarities
between
the IIA and IIB theories on $X$
were  studied some time ago
\cite{CFG:II}, and the connection to mirror symmetry was pointed
out  in \cite{dWvP:quat,Str:con,udual}.}
(We shall refer to these
compactified theories as IIA$_X$ and IIB$_Y$, respectively.)
One of the remarkable properties of mirror symmetry
is the relationship which it establishes between the integer cohomology
group $H^3(X,\Z)$ and the ``vertical'' integer cohomology
$\bigoplus H^{2k}(Y,\Z)$ of the mirror
partner \cite{AL:qag,mirrorguide,2param1,predictions}.
This property is somewhat mysterious from the point of view of conformal
field theory, since the integer cohomology plays no apparent r\^ole
there.

In string theory, the integer cohomology groups $H^3(X,\Z)$ and
$\bigoplus H^{2k}(Y,\Z)$  find their
proper r\^ole as likely candidates for describing
the set of discrete shifts in the massless Ramond-Ramond scalars
of the IIA$_X$ and IIB$_Y$ theories.  The equivalence between the two can
then be seen as a first step in establishing the equivalence between
the full IIA$_X$ and IIB$_Y$ theories.  However, there are some subtleties
in the equivalence between integer cohomology groups which will lead us to
the conclusion that the lattice of discrete shifts in Ramond-Ramond fields
for a type IIB theory depends on the $B$-field as well as on
$\bigoplus H^{2k}(Y,\Z)$.  This is somewhat reminiscent of the theta-angle
dependence which occurs
in Witten's discussion of charge quantization of dyons
\cite{Wit:dyons}.

\section{MIRROR SYMMETRY IN STRING THEORY}

Mirror symmetry  \cite{dixon,LVW,CLS,GP}
was originally formulated as a property of
two-dimensional nonlinear
sigma-models on Calabi--Yau manifolds.  These sigma-models flow to
$N=(2,2)$ superconformal field theories in the infrared, and it is
possible to find ``mirror pairs'' of such manifolds whose associated CFTs
become
isomorphic once the sign of the left-moving $U(1)$-charge has been
changed in one of the two theories.
Mirror symmetry
relates
the CFT moduli spaces of the two Calabi--Yau manifolds, producing
local isomorphisms
between the K\"ahler moduli space of one manifold and the
complex-structure moduli space of its mirror partner.

If a Calabi--Yau manifold $X$ of dimension $d$ is used to compactify
the type IIA or IIB string, the $N=(2,2)$ moduli space of $X$
is embedded in the NS-NS sector of the moduli space of
the effective $(10{-}d)$-dimensional
theory.\footnote{The conformal field theory on $X$ can also be
regarded as an $N=(0,2)$ theory and used to compactify the heterotic
string, but as the implications of mirror symmetry are not
as well-understood in this context
we will focus on the type II string.}
To see how a sign change in a worldsheet $U(1)$-charge would
 affect  the effective field theory,
we consider
Minkowski space $\R^{9-d,1}$, and work in
light-cone
gauge in which the spacetime fermions transform in spinor
representations of
$SO(8-d)$.
The left-moving $\hat{u}(1)$ affine algebra from the $N=(2,2)$ algebra
of the superconformal field theory
lies in the affine algebra $\widehat{so}(8-d)$,
and the weights of $SO(8-d)$ are charged under $U(1)$.
A change of sign in left-moving worldsheet $U(1)$-charge must therefore
be accompanied by an action of the automorphism $C$
 of the weight space of $SO(8-d)$ which
changes the signs of all of the weights.

If $d$ is divisible by $4$, then $C$ maps each spinor representation
of $SO(8-d)$ to itself.  Thus, the worldsheet sign-change will leave
the IIA theory as a IIA theory, and the IIB theory as a IIB theory.
On the other hand, if $d$ is congruent to $2$ modulo $4$, then $C$
maps each spinor to the spinor of opposite chirality.  It follows
that in this case, an exchange between the IIA and IIB theories must
accompany the worldsheet sign-change.

If $X$ and $Y$ are a mirror pair with $d$
congruent to $2$ modulo $4$, then mirror symmetry should relate IIA$_X$
to IIB$_Y$ and vice versa.
This equivalence can be considered as an analogue of the supersymmetric
$R{\leftrightarrow}1/R$
equivalence \cite{DHS:IIAB,DLP:IIAB}, which identifies
IIA$_{S^1}$ at large radius with IIB$_{S^1}$ at small radius;
there are extensions of this to compactifications on a torus of arbitrary
dimension.
In the case of $d=2$, we have
$X=Y=T^2$ and mirror symmetry can be derived from this
$R{\leftrightarrow}1/R$ equivalence.
In the case of $d=6$, however, mirror
symmetry should provide a {\em further}\/ equivalence between type II theories
which is not a direct consequence of $R{\leftrightarrow}1/R$,
since most Calabi--Yau threefolds
do not have an $S^1$ factor or even an action of $S^1$.\footnote{There
has been some recent speculation \cite{PT}
 about yet a third type of equivalence between
type II theories, which would relate IIA$_{J^7}$ to IIB$_{J^7}$ for
compactifications on a Joyce manifold $J^7$ of holonomy $G_2$ \cite{joyce12}.}

On the other hand,
if $X$ and $Y$ are a mirror pair and $d$ is divisible by $4$,
mirror
symmetry relates IIA$_X$ to IIA$_Y$ and IIB$_X$ to IIB$_Y$.  The
familiar cases of mirror symmetry in these dimensions are $X=Y=T^4$
and $X=Y=K3$, both with $d=4$.
Because these
manifolds are self-mirror,
mirror symmetry acts as a discrete identification on the moduli spaces.
(In the $K3$ case, this has been used to establish \cite{stringK3}
 that the discrete
identifications which act on the moduli space for IIA$_{K3}$
are precisely the same as those of the moduli space for
the heterotic string compactified on $T^4$ \cite{N:torus,HNW:torus}, as had
been
conjectured by Seiberg \cite{Sei:K3}.)
There may well be similar mirror identifications in $d=8$, compactifying
the type II string
on one of Joyce's manifolds \cite{joyce3}
with holonomy $Spin(7)$.\footnote{Some preliminary suggestions about
mirror phenomena in this case
were made in \cite{SV}.}

\section{THE CONFORMAL FIELD THEORY MODULI SPACE} \label{sect:cft}

When passing from a classical to a quantum moduli space, new degrees
of freedom may arise from the following mechanism.  A continuous symmetry
of the classical theory may be broken to a discrete symmetry of
the quantum theory.  There will be some massless scalar
fields whose expectation values
in the classical theory can be shifted to some fixed value (typically to
zero) by exploiting
the symmetry.  In the quantum theory, however,
the possible shifts are restricted to
a discrete set, and the expectation values of these fields (modulo the
discrete identifications) provide the new degrees of freedom.  We shall
use the term ``semiclassical moduli space'' to refer to the space
obtained from the classical moduli space by including these new
degrees of freedom.  The discrete identifications are a quantum effect
which must be respected by any further perturbative or nonperturbative
quantum corrections to the moduli space.

A familiar example of this mechanism is provided by
the nonlinear sigma-model on
a Calabi--Yau manifold $X$.  This model
behaves at large radius like a field theory on
$X$, and the ``classical'' moduli space is the space of possible
Ricci-flat metrics (normalized to have a fixed volume).  The volume
$V$, or radius $R=V^{1/d}$, of the manifold measures the size
of quantum effects in the theory (which are suppressed at large radius).
The new degree of freedom which arises in the quantum
moduli space is the so-called $B$-field,
which is a harmonic $2$-form on $X$.\footnote{We ignore the additional
degree of freedom which is provided by torsion in $H_2(X)$
\cite{SW:spin,Vafa:tor,DW:gp,chiral}.}
It is well-defined only up to shifts $B\mapsto B+\delta B$,
where $\delta B$ is a $2$-form which represents a class in {\em integer}\/
cohomology, i.e., $\delta B\in H^2(X,\Z)$.  The fact that it is
the integer cohomology which describes the discrete shifts follows
from the presence of a ``topological'' term $S_{top}=\int_\Sigma \varphi^*(B)$
in the sigma-model action. (Here $\varphi$ is a map from the worldsheet
$\Sigma$ to $X$).  If the action is normalized so that
$\exp(2\pi i\,S_{top})$ is what appears in physically measurable quantities,
then the discrete symmetry of $B$ must be represented by integer cohomology
in order that  $\int_\Sigma\varphi^*(\delta B)$ will be an integer, and
hence that $\exp(2\pi i\,\int_\Sigma\varphi^*(\delta B))$
will be equal to $1$.

The semiclassical description of the CFT moduli space is only valid at large
radius, and indeed the structure of the moduli space at small radius
is known to be substantially altered by nonperturbative effects
(worldsheet instantons).  Mirror
symmetry (in conjunction with certain non-renormalization theorems
\cite{DG:exact})
has been very useful in understanding the structure of the moduli
space in these regions \cite{catp},
since it relates regions of small radius on $X$
to regions of large radius on the mirror partner $Y$.  More precisely,
the condition that the mirror partner $Y$ be at large radius translates
into a requirement that the complex structure on $X$ lie in a certain
region of the complex-structure moduli space (with no condition on
the radius).  The complex-structure moduli space is unaffected by
the worldsheet instantons.

Consider now a path in the moduli space of $Y$ in which the metric is fixed,
and the $B$ field takes the value $B_0+t\,\delta B$, for $0\le t\le1$.
Thanks to the discrete identification, this path forms a loop in
the semiclassical moduli space.  The mirror image of such a loop
is a loop in the complex-structure moduli space $\cM_X$ of $X$ which encircles
a boundary component of that moduli space.
(Transporting structures along such loops will play an important part in
our analysis below.)
As the radius of $Y$ increases,
the corresponding loop $\gamma$
in $\cM_X$ shrinks towards the boundary component.
In fact, if we rescale the metric on $Y$ by $g_{ij}\mapsto\lambda g_{ij}$
and let
$\lambda\to\infty$, then the loops $\gamma(\lambda)$ will sweep out a
punctured disk whose limit point is on the boundary component
(cf.\ \cite{where}).  For an appropriate compactification $\overline{\cM_X}$
of the complex-structure moduli space $\cM_X$,
the mirror of a large radius limit
point appears as an intersection of $r=\dim(\cM_X)$ boundary components,
and the disks swept out by $\gamma_1(\lambda), \dots,
\gamma_r(\lambda)$ provide local complex coordinates $t_1, \dots, t_r$ on
the compactification.

\section{THE TYPE IIA MODULI SPACE}

We specialize now to the case $d=6$, so that $X$ is a Calabi--Yau threefold.
We assume that the first Betti number of $X$ is $0$, or equivalently,
that Ricci-flat metrics on $X$ have holonomy precisely $SU(3)$.
If we compactify the IIA string on $X$,
the massless scalar spectrum of the theory consists of the metric, the
$B$-field, the axion $\theta$ and the
dilaton $\phi$
in the NS-NS sector, and a field which corresponds to
a harmonic $3$-form $C$ in the R-R
sector.\footnote{We again ignore any effects that may be associated with
torsion in homology.}
The ``classical'' moduli space for the IIA$_X$ theory coincides with the
conformal field theory moduli space described above---at large radius,
it is accurately described by the metric and $B$-field, modulo discrete
shifts of the $B$-field.
The dilaton measures the size of quantum effects
in the IIA theory, and the other new fields---the
axion and the R-R $3$-form $C$---should be subject to discrete symmetries
as we have discussed.  In the case of the axion, this is well-known:
the shift is by integer multiples of $2\pi$, and
it is common to use a parameter
$\exp(8\pi^2S)=\exp(-i\theta+8\pi^2e^{-2\phi})$
which combines the axion and dilaton and also
implements the identifications by discrete
shifts.
However, in the case of the $3$-form, the precise nature of the
discrete shift is more difficult to pin down.

If an action for $2$-branes plays a r\^ole in the eventual nonperturbative
formulation of type IIA string theory, then a topological term in that
action of the
form $S_{top}=\int_M\psi^*(C)$ would cause the discrete shift to take the
form $C\mapsto C+\delta C$ with $\delta C\in H^3(X,\Z)$.
(Here $\psi$ is a map from the world-volume $M$ to the target space $X$,
and the argument is completely analogous to the case of the $B$-field.)
 However, it would
be preferable to arrive at this conclusion without assuming such details about
the form of the nonperturbative theory.
In \cite{udual} it was argued that for many Calabi--Yau threefolds, mirror
symmetry combined with $S$-duality for the type IIB theory implies that the
discrete shifts $\delta C$ must fill out a finite index subgroup of
$H^3(X,\Z)$.

We shall assume (for simplicity)
that $H^3(X,\Z)$ provides the correct set of discrete
shifts for the quantum theory.\footnote{The bulk of our analysis could
be restated (in a more cumbersome fashion) for the case of a finite
index subgroup.}  We shall also assume that $X$ has a mirror partner
$Y$.
Then the semiclassical moduli space of the IIA$_X$ theory has the
following description.  The CFT moduli space is essentially a product
of the complex-structure moduli spaces $\cM_X$ and $\cM_Y$ of $X$
and $Y$.  (There are some subtleties about that statement, but they
need not concern us here.)  The R-R $3$-form $C$ will transform in
a vector bundle $\cE_X\to\cM_X$ whose fibers are the cohomology
groups $H^3(X,\R)$.  Within that bundle is a bundle of lattices
$\cE_X^{\Z}$ which describe the discrete identifications
(the fibers of which are $H^3(X,\Z)$).
The complex-structure moduli and $C$ together fill out the quotient
$\cE_X/\cE_X^{\Z}$,
a bundle of tori.  Finally, the axion-dilaton field $\exp(8\pi^2S)$ transforms
in a $\C^*$-bundle $\cL^*\to(\cE_X/\cE_X^{\Z})$, and the entire semiclassical
moduli space can be  described as $\cL^*\times\cM_Y$
(up to the subtlety about the product structure alluded to earlier).

The bundle of tori $(\cE_X/\cE_X^{\Z})\to\cM_X$ is a familiar object
in algebraic geometry, called the family of {\em intermediate Jacobians}.
These tori come equipped with natural complex structures by means of
the isomorphisms
\begin{equation}
H^3(X,\R)\cong H^{3,0}(X_t)\oplus H^{2,1}(X_t)
\end{equation}
for $t\in\cM_X$.
Griffiths \cite{griffiths} proved that these complex
structures vary holomorphically, that is, the total space
$\cE_X/\cE_X^{\Z}$ is itself a complex manifold and the map
to $\cM_X$ is  holomorphic.
Donagi and Markman
\cite{DM:cubics}
have recently shown that $\cE_X/\cE_X^{\Z}$ has the additional structure
of being a complex contact manifold, and that the $\C^*$-bundle
$\widetilde{\cL}^*\to\cE_X/\cE_X^{\Z}$ whose fibers are the nonzero
elements of $H^{3,0}(X_t)$ is the associated complex symplectic manifold.
This is precisely the geometry that one expects will underlie a
hyper-K\"ahler metric.  (The non-compactness of $\widetilde{\cL}^*$
prevents us from immediately concluding that such a metric exists.)
It is tempting to identify $\widetilde{\cL}^*$ with
$\cL^*$ since a quaternionic K\"ahler metric is expected on the latter,
of which the putative hyper-K\"ahler metric is perhaps a limit.  It should
be possible to settle this question and
determine the precise nature of the metric on $\cL^*$
by means of the explicit ``c-map'' of \cite{FS:quat}.  This issue is currently
under investigation.

We can expect nonperturbative corrections to the semiclassical moduli
space  of various
kinds.  A mirror version of the conifold transitions of \cite{GMS} should
link together some (perhaps all) of these moduli spaces.  Other nonperturbative
effects should modify the structure of this space at strong coupling
\cite{BBS}.

\section{MONODROMY}

The description we have given of the semiclassical type IIA moduli space
 was complicated by the unavoidable fact that the
bundle of lattices $\cE_X^{\Z}$ is {\em not}\/ trivial over $\cM_X$.
If we follow the $H^3(X_t,\Z)$ lattice as $t$ traverses a loop
$\gamma$ in $\cM_X$, then the lattice undergoes a {\em monodromy
transformation}\/ represented by a matrix $T_{\gamma}$.
In particular, it is not usually possible to find a single-valued
function $t\mapsto\Gamma(t)\in H^3(X_t,\Z)$ as $t$ traverses such a loop.

The behavior
of these monodromy matrices is not arbitrary, however.  By the
monodromy theorem \cite{monodromy}, all eigenvalues of $T_\gamma$
are roots of unity.  Furthermore, if we work near the mirror of a
large radius limit point, using loops $\gamma_1, \dots, \gamma_r$
as at the end of section \ref{sect:cft}, then the eigenvalues are
all $1$ and the logarithms $N_j:=\log T_{\gamma_j}$ are a commuting
set of nilpotent
matrices.
The nilpotent orbit theorem \cite{schmid} says that although
a function of the form $t\mapsto\Gamma(t)$ taking values in $H^3(X_t,\Z)$
is not single-valued, the function
\begin{equation}
t\mapsto\exp\left(-\sum \frac{\log t_j}{2\pi i}\,
N_j\right)\,\Gamma(t)
\end{equation}
{\em is}\/ single-valued and behaves well at the boundary,
where $t_1, \dots, t_r$ are the coordinates
associated to $\gamma_1, \dots, \gamma_r$.

(We wish to stress that this same result would have been obtained for
{\em any}\/ lattice within $H^3(X_t,\R)$; it does not depend on our
assumption that the discrete shifts correspond to integer cohomology.)

The mirror of the monodromy transformations $\exp(N_j)$ have a very
natural description.  Traversing the loop $\gamma_j$ in $\cM_X$
corresponds to following a path $B= B_0+t\,(\delta B)_j$,
$0\le t\le1$ in the NS-NS moduli space of $Y$, where
$(\delta B)_j\in H^2(Y,\Z)$.  The action of the monodromy
on $H^3(X,\R)$ is then mapped to an action of the  $B$-field shift
$(\delta B)_j$  on the vertical cohomology $H^{even}(Y,\R)$ described
by
\begin{equation}\label{mapping}
C\mapsto \exp(\,(\delta B)_j) \wedge C \in H^{even}(Y,\R),
\end{equation}
where we write $C=\sum_{k=0}^3 (3-k)!\,C_{2k}$, and
$\exp(B)=1+B+\frac12(B\wedge B)+\frac16(B\wedge B\wedge B)$, both
regarded as elements of $H^{even}(Y,\R)$.  The factor of $(3-k)!$
is designed to make the integer cohomology work out appropriately,
and indeed if $\delta B\in H^2(Y,\Z)$ then $\exp(\delta B)$
will map $\bigoplus_{k=1}^3 (3-k)!\,H^{2k}(Y,\Z)$ to itself.

In all known examples of mirror pairs for
 which the integer cohomology has been computed,
that cohomology exhibits the following beautiful structure
(see \cite{predictions} for a review):
$H^3(X,\Z)$ is mapped to $H^{even}(Y,\Z)$ in such a way that
the monodromies $\exp(N_j)$ are sent to the mappings of eq.~(\ref{mapping}).
(Both of these transformations are well-defined on integer cohomology.)
It is conjectured that this is always the case.

\section{THE TYPE IIB MODULI SPACE}

We continue to let $Y$ denote
 a Calabi--Yau threefold with $b_1(Y)=0$ which has $X$ as its mirror
partner.
The semiclassical moduli space for the type IIB$_Y$ theory is more
difficult to describe than that for the type IIA theory, in part because we
lack a Lagrangian formulation for type IIB theories.  The NS-NS sector
of the massless spectrum still
 consists of the metric, the
$B$-field, the axion and the
dilaton; however, the R-R sector is harder to identify.
The R-R field content of the IIB theory in 10 dimensions consists of
a $0$-form, a $2$-form and a self-dual $4$-form, which one might expect
to give rise by dimensional reduction
to two harmonic $0$-forms, two harmonic $2$-forms and a
harmonic $4$-form.
  However, the self-duality condition on the
field strength of the $4$-form reduces the R-R degrees of freedom to
two $0$-forms, one $2$-form and one $4$-form.
We shall replace one
of the $0$-forms by its Hodge star (a $6$-form),
leaving us with R-R fields
$C_{2k}$: a harmonic $2k$-form for each of $k=0, 1, 2, 3$.  This
replacement is motivated
by the way in which mirror symmetry identifies $H^3(X)$ with $H^{even}(Y)$.
In the absence of a Lagrangian our identification of the R-R
fields is somewhat tentative, but we have at least gotten the
number of degrees of freedom right.

The semiclassical moduli space for the type IIB$_Y$ theory can now
be built up as before.  We start with the CFT moduli space
$\cM_X\times \cM_Y$, include the axion-dilaton
 field as transforming
in a $\C^*$-bundle $\cL^*$, and describe the R-R fields as taking
values in a bundle over the NS-NS moduli space
with fibers $H^{even}(Y,\R)/\Lambda$, where
$\Lambda$ represents the discrete shifts.
As in the type IIA case, we expect nonperturbative corrections---both
at weak coupling due to conifold transitions \cite{GMS}, and at strong
coupling.

The semiclassical moduli spaces of IIA$_X$ and IIB$_Y$ cannot be
globally isomorphic, for a simple reason: the R-R fields in the IIB
case are modeled on $H^{even}(Y)$, which in conformal field theory
was subject to worldsheet instanton corrections, whereas the R-R
fields in IIA are modeled on $H^3(X)$ which has no such corrections.
It is expected that the worldsheet instanton corrections will be
mimicked by nonperturbative solitons in the IIB theory \cite{BBS}
which will correct the moduli space at strong coupling.
In any event, any proposed mirror isomorphism between IIA$_X$ and IIB$_Y$
semiclassical
moduli spaces should only be considered locally, near weak coupling.
We can still reliably study the discrete shifts there.

We have not yet determined the discrete identifications $\Lambda$
of the R-R fields.
At first glance it would appear that the most natural guess for this
lattice
 would be the vertical integer cohomology
$H^{even}(Y,\Z)$.  If mirror symmetry is to hold, however, this
guess cannot be correct.  For if it were,
 the semiclassical
moduli space would be a {\em trivial}\/ bundle over the NS-NS moduli space,
with fiber $H^{even}(Y,\R)/H^{even}(Y,\Z)$.  That is not the structure
we found on the type IIA side, since the monodromy is missing.
The only alternative is that
{\em the precise values of the possible discrete shifts
$\delta C_{2k}$  must depend on the value of $B$}!

As we saw in the previous section, upon shifting the $B$-field by
$\delta B\in H^2(Y,\Z)$, the vertical integer cohomology classes
will shift by  wedging with $\exp(\delta B)$.
The most straightforward way to reproduce this
monodromy behavior
is to postulate the following structure for the discrete
shifts.\footnote{In \cite{udual}, we described this structure in terms of the
quantum cohomology ring of $Y$.  The two formulations are essentially
equivalent.}
Let $C=\sum_{k=0}^3 (3-k)!\,C_{2k}$, and let $\delta C$ be the
discrete shift.
The condition we should require is:
\begin{equation}
\exp(B)\wedge (\delta C) \in \bigoplus_{k=0}^3\, (k!)H^{6-2k}(Y,\Z).
\end{equation}
It is worthwhile to write this out more explicitly:
{\arraycolsep=1pt
\begin{eqnarray} \label{shifts}
\delta C_0&\in H^0(Y,\Z)& \nonumber \\
\delta C_2+3B\wedge \delta C_0&\in H^2(Y,\Z)& \nonumber \\
\delta C_4+2B\wedge \delta C_2+3B\wedge B\wedge \delta C_0&\in H^4(Y,\Z)&\\
\delta C_6+B\wedge \delta C_4+B\wedge B\wedge \delta C_2 \quad && \nonumber \\
{}+B\wedge B\wedge B\wedge \delta C_0&\in H^6(Y,\Z).& \nonumber
\end{eqnarray}
Our condition has the property that
when we shift $B$ to $B+\delta B$ then the condition changes by
wedging with $\exp(\delta B)$, precisely reproducing the anticipated
monodromy effect.  Of course it is also possible to imagine a more complicated,
nonlinear $B$-field dependence with the same property.
}

What kind of nonperturbative effect in type IIB theory could produce
such a condition?  We would appear to need topological terms in
actions for $p$-branes for each of $p=-1,1,3,5$.  For example,
the third condition might follow from a $3$-brane action with
a topological term of the form
\begin{equation}
S_{top}=\int_{M^4} \eta^*(C_4+2B\wedge C_2 + 3B\wedge B\wedge C_0)
\end{equation}
($\eta$ being a map from the worldvolume to the target space).
Why the term should take precisely this form is somewhat mysterious.

It is tempting to speculate that a related $B$-field dependence will
appear in the charge quantization rules for R-R gauge fields,
similar to Witten's discussion of charge quantization for dyons
in the presence of a theta-angle \cite{Wit:dyons}.

\section{CONCLUSIONS}

Let us summarize our analysis in the following way.  There are three
conjectures which seem very reasonable, and which reinforce each
other nicely.  The first is that the discrete shifts of the R-R
fields in the IIA$_X$ theory are given by $H^3(X,\Z)$ when
$b_1(X)=0$.\footnote{If $b_1(X)\ne0$, we should actually expect
some dependence on the $B$-field in this case as well, through
terms such as $B\wedge C_1$.  This follows from the analysis of
$K3\times T^2$ given in \cite{udual}.}  The second is that the
integer cohomology and integer monodromies are preserved by mirror
symmetry.  (There is concrete evidence for this second conjecture.)
And the third is that the discrete shifts of the R-R fields
in the IIB$_Y$ theory are given by eq.~(\ref{shifts}).  Together,
these conjectures suggest a coherent picture of mirror symmetry
in string theory, giving a local isomorphism between
  the semiclassical moduli spaces at weak coupling.  It is to be
hoped that the isomorphism will extend to the full moduli spaces
once nonperturbative effects are taken into account.

\section*{ACKNOWLEDGEMENTS}

It is a pleasure to thank Paul Aspinwall for collaboration on some of
the work presented here.  I would also like to thank Jacques Distler,
Brian Greene, Shamit Kachru, Vadim Kaplunovsky, Ronen Plesser and
Andy Strominger for discussions.  This work was supported in part by
National Science Foundation grant DMS-9401447.

\ifx\undefined\bysame
\newcommand{\bysame}{\leavevmode\hbox to3em{\hrulefill}\,}
\fi

\end{document}